\documentclass[twocolumn,showpacs,preprintnumbers,amsmath,amssymb]{revtex4}

\usepackage{graphicx}
\usepackage{dcolumn}
\usepackage{bm}
\def\vector#1{{\boldsymbol{#1}}}

\def\vk{{\vector k}}

\def\vr{{\vector r}}
\def\vR{{\vector R}}

\def\vF{{v_{\it F}}}

\def\Tc{{T_{\it c}}}

\def\BETSFeCl{\mbox{${\rm (BETS)_2FeCl_4}$}}

\def\RSGCO{\mbox{${\rm RuSr_2GdCu_2O_8}$}}

\def\hsp#1{\hspace{#1ex}}

\def\Tc{{T_{\it c}}}

\def\lsim{\stackrel{{\textstyle<}}{\raisebox{-.75ex}{$\sim$}}}
\def\gsim{\stackrel{{\textstyle>}}{\raisebox{-.75ex}{$\sim$}}}

\def\eq.#1{eq.~(\ref{#1})}
\def\refeq#1{(\ref{#1})}

\newcommand\Equation[2]{
\begin{equation}\label{#1}
#2
\end{equation}
}

\begin{document}

\title{
Reduction of Pauli paramagnetic pair-breaking effect \\
in antiferromagnetic superconductors
}


\author{Hiroshi Shimahara}



\affiliation{
Department of Quantum Matter Science, ADSM, Hiroshima University, 
Higashi-Hiroshima 739-8530, Japan
}


\date{Received June 17, 2004}

\begin{abstract}
Antiferromagnetic superconductors in a magnetic field are studied. 
We examine a mechanism which significantly reduces 
the Pauli paramagnetic pair-breaking effect. 
The mechanism is realized even in the presence of 
the orbital pair-breaking effect. 
We illustrate it using a three-dimensional model 
with an intercalated magnetic subsystem. 
The upper critical field is calculated for various parameters. 
It is shown that the upper critical field can reach several times 
the pure Pauli paramagnetic limit. 
The possible relevance to the large upper critical field observed in 
the heavy fermion antiferromagnetic superconductor ${\rm CePt_3Si}$ 
discovered recently is briefly discussed. 
We try to understand 
the large upper critical field in the compound ${\rm CePt_3Si}$ 
and field-induced superconductivity in the compound ${\rm CePb_3}$ 
within a unified framework. 
\end{abstract}


\maketitle


In this paper, we examine superconductivity 
in the presence of a coexisting antiferromagnetic long-range order. 
In a magnetic field, 
some types of antiferromagnet exhibit canted spin structures, 
which give rise to a net ferromagnetic moment 
proportional to the magnitude of the magnetic field. 
The ferromagnetic moment creates an exchange field 
on the conduction electrons by the Kondo interaction. 
When the Kondo interaction is antiferromagnetic, 
the exchange field reduces the effective Zeeman energy. 
As a result, 
the Pauli paramagnetic pair-breaking effect~\cite{Cha62,Clo62} 
can be significantly reduced~\cite{Shi02b,Shi03}, 
and the upper critical field $H_{\it c2}$ can largely exceed 
the Pauli paramagnetic limit $H_{\it P}$ 
(Chandrasekhar and Clogston limit).

The analogous compensation mechanism 
in superconductors with uniformly aligned spins is known as 
the Jaccarino-Peter mechanism~\cite{Jac62}, 
which explains the field-induced superconductivity (FISC) 
in the compounds ${\rm Eu_xSn_{1-x}Mo_6S_8}$~\cite{Meu84}, 
${\rm CePb_3}$~\cite{Lin85}, 
and $\lambda$-\BETSFeCl~\cite{Uji01,Cep02,Shi02a,Hou02}. 
However, the resultant phase diagrams of the present mechanism 
are quite different from that of the Jaccarino-Peter mechanism. 
The superconductivity occupies a large single area 
including the zero field ($H = 0$) in the $T$-$H$ phase diagram, 
while depending on the parameter values, 
the FISC may occur, 
where $T$ and $H$ denote the temperature and the magnetic field, 
respectively. 
In our previous papers~\cite{Shi02b,Shi03}, 
we have proposed this mechanism in a multilayer system 
in a parallel magnetic field. 
We have adopted such a system, 
because the reduction mechanism is clearly illustrated. 
In this paper, we extend the same mechanism to more general systems 
in the presence of the orbital pair-breaking effect. 
The magnitude and the temperature dependence 
of the resultant upper critical field can be similar to 
those of the spin-triplet superconductivity 
of equal spin pairing.

The present work was motivated by 
the heavy fermion antiferromagnetic superconductor 
${\rm CePt_3Si}$ discovered recently~\cite{Bau04}. 
Due to the heavy quasi-particle mass, 
the orbital pair-breaking effect is much weaker than 
in conventional metals. 
Furthermore, the compound exhibits antiferromagnetic transition 
at $T_{\it AF} \approx 2.2\,{\rm K}$ 
and superconducting transition 
at $T_{\it c}^{(0)} \approx 0.75\,{\rm K}$ at the zero field. 
The upper critical field $H_{\it c2}(0) \approx 5\,{\rm T}$ at $T = 0$ 
is much larger than the Pauli paramagnetic limit 
estimated by the simplified formula 
$H_{\it P} \approx 1.86 [{\rm T/K}] \times T_{\it c}^{(0)} [{\rm K}] 
\approx 1.4 \,{\rm T}$. 
Such a large $H_{\it c2}$ seems to suggest spin-triplet pairing, 
although the lack of inversion symmetry of this compound seems 
disadvantageous to spin-triplet pairing. 
Recently, 
Frigeri {\it et al.} have shown that spin-triplet pairing 
is not entirely excluded by the lack of inversion symmetry. 
Therefore, equal spin pairing may be 
the reason for a large $H_{\it c2}$~\cite{Fri04}. 
In this paper, however, we propose another scenario based on 
antiparallel spin pairing including spin-singlet pairing. 
Samokhin {\it et al.} have carried out band structure calculations 
and proved within their theory that 
the order parameter must be an odd function of 
the momentum $\vk$~\cite{Sam04}. 
We should note that this does not mean the occurrence of 
equal spin pairing~\cite{Sam04Note}. 
Hence, our scenario does not contradict their result.

In principle, we might be able to test these scenarios 
by rotating the magnetic field. 
The upper critical field $H_{\it c2}$ must be highly anisotropic, 
if the large $H_{\it c2}$ is due to equal spin pairing 
and the ${\vector d}$-vector is fixed. 
For example, in the compound ${\rm CePt_3Si}$, 
the strong spin-orbit interaction favors 
a particular ${\vector d}$-vector~\cite{Fri04}. 
For the magnetic field parallel to the ${\vector d}$-vector, 
the upper critical field must be strongly suppressed. 
In contrast, 
the upper critical field can be nearly isotropic 
in the present mechanism. 
The Knight shift measurement might also give 
useful informations~\cite{Sam04Note}.

As mentioned above, the purpose of this study is 
to examine the present mechanism 
in the presence of the orbital pair-breaking effect. 
Therefore, we examine a three-dimensional system 
with an intercalated magnetic subsystem. 
We divide the magnetic subsystem into two sublattices, 
which we call A and B. 
We approximate the conduction electron system with 
an isotropic continuum system for simplicity.

When we apply the theory to the compound ${\rm CePt_3Si}$, 
the A and B sublattices are alternate layers of Ce atoms, 
and the spins on each layer are 
ferromagnetically ordered~\cite{Met04}. 
The superconductivity and the antiferromagnetism may exist 
in the same degrees of freedom in this compound, 
but we simplify the situation 
by dividing the degrees of freedom into two coupled systems. 
We also ignore the spin-orbit interaction 
and the lack of inversion symmetry for simplicity, 
because they do not affect the present mechanism.

First, let us describe the spin structure in the magnetic subsystem. 
We consider the situation in which 
the antiferromagnetic transition occurs at a temperature 
much higher than the superconducting transition temperature, 
which is consistent with the observation 
in the compound ${\rm CePt_3Si}$~\cite{Bau04}. 
Since the ordered state is rigid, 
we neglect the modification of the magnetic structure 
by the occurrence of superconductivity. 
Therefore, 
we consider the cases 
in which the magnetic subsystem is effectively described 
by the Hamiltonian 
\Equation{eq:Hs}
{
     H_{s} = \sum_{i,j}
                J_{ij} \, 
                   {\vector S}_{i} \cdot {\vector S}_{j} 
                 + {\sum_{i}}
                     g_{s} \mu_{B} {\vector H} 
                     \cdot {\vector S}_{i} , 
     }
where $g_{s}$ and $\mu_{B}$ 
denote the $g$-factor of the localized spins 
and the Bohr magneton~\cite{NoteHs}. 
We define $z$ as the number of antiferromagnetic bonds 
for a given site. 
We set $J_{ij} = 0$ 
except for the nearest neighbor sites $(i,j)$. 
For the compound ${\rm CePt_3Si}$, 
we assume that $J_{ij} = J_{\parallel} < 0$ 
for $(i,j)$ on the same layer (the same sublattice), 
while $J_{ij} = J_{\perp} > 0$ 
for $(i,j)$ on the adjacent layers (the different sublattices). 
In this case, we have $z = 2$. 
We could examine other crystal structures. 
For example, we have $z = 6$ in the system on the cubic lattice. 
In any case, we assume that $J_{ij} = J > 0$ 
for the antiferromagnetic bonds.

We take the $x$-axis of the spin space as being 
in the direction antiparallel to the external field. 
Hence, we set ${\vector H} = ( - H, 0, 0 )$. 
The directions of the axes of the spin space 
do not necessarily coincide with the crystal axes 
in the present model. 
For $T \lsim T_{c}^{(0)} \ll T_{\it AF}$, 
we neglect the thermal fluctuations 
and regard the localized spins as classical variables. 
If we ignore the anisotropy energy of the localized spins 
in the magnetic layers, 
the sublattice magnetization appears in the direction perpendicular 
to the applied field, 
since it is energetically favored. 
Hence, we introduce the classical variable $\theta$ by 
\Equation{eq:Siclassical}
{
     {\vector S}_{i} = {\bar S} \, ( \sin \theta , 0 , \pm \cos \theta ) , 
     }
where we define the $z$-axis of the spin space 
as being in the direction of the sublattice magnetization. 
For the double sign $\pm$ in \eq.{eq:Siclassical}, 
we adopt $+$ and $-$ signs 
when site $i$ belongs to the A and B sublattices, 
respectively. 
Here, we have introduced the magnitude of the localized spins ${\bar S}$ 
taking into account the shrinkage of spins by fluctuations. 
The total energy of the magnetic subsystem is expressed as 
\Equation{eq:energylocalizedspins}
{
     E(\theta) = N {\bar S} \, 
          \Bigl [ 
            - z J {\bar S} \cos 2 \theta 
            - 2 h_{\it s} \sin \theta 
          \Bigr ] 
            + E_0 
          , 
     }
where $E_0$ is a constant and $h_{\it s} \equiv g_{\it s} \mu_{\it B} H/2$. 
The total energy becomes minimum when 
\Equation{eq:sinthetaH}
{
     z J {\bar S} \sin \theta = h_{\it s} 
     }
for $h_{\it s}/zJ{\bar S} \leq 1$, {\it i.e.}, 
$H \leq 2 zJ{\bar S}/ g_{\it s} \mu_{\it B} \equiv H_{\it AF}$. 
For a high field in which $H \geq H_{\it AF}$, 
the staggered moment disappears ($\theta = \pi/2$). 
From now on, we mainly consider a weak field $H \leq H_{\it AF}$, 
where the antiferromagnetic order persists. 
The localized spins are also described by the spin density 
\Equation{eq:spindensity}
{
     {\vector S}(\vr) 
       = \sum_{i} \delta^{(3)}(\vr - \vR_i) \, {\vector S}_i , 
     }
where $\vR_i$ denotes the position of site $i$.

We treat the degrees of freedom of the conduction electrons 
as a continuum model. 
The net spin moment ${\bar S} \sin \theta$ gives rise to the exchange field 
on the conduction electrons. 
In order to reproduce the exchange field, 
we employ an extended Kondo Hamiltonian defined as 
\Equation{eq:HK}
{
     \begin{split}
     H_{\it K} = & 
               \int d^3 \vr \int d^3 \vr' \, 
                   {\hat J}_{\it K}(\vr - \vr') \, 
             \\
             & \times 
                 {\vector S}(\vr) 
                 \cdot 
               \sum_{\sigma  \sigma'}
               \bigl [ 
                 \psi_{\sigma}^{\dagger}(\vr') 
                   {\boldsymbol{\sigma}}_{\sigma \sigma'} 
                 \psi_{\sigma'}(\vr') 
               \bigr ] , 
     \end{split}
     }
where ${\boldsymbol{\sigma}}_{\sigma \sigma'}$ 
and $\psi_{\sigma}(\vr)$ 
denote the Pauli matrices and 
the annihilation operator of the conduction electron at $\vr$. 
Here, we note that \eq.{eq:HK} is used in the calculation of 
the order parameter $\Delta(\vr)$, 
which appreciably varies in the length scale of 
the coherence length $\xi \gg a$, where $a$ denotes the lattice constant. 
Therefore, after inserting \eq.{eq:spindensity} in \eq.{eq:HK}, 
we replace the spin variable ${\vector S}_i$ 
by the spatial average ${\langle {\vector S} \rangle}_{\xi i}$ 
over the volume $v_{\xi i}$ on the order of $\xi^3$ near site $i$, 
which is defined as 
${
     {\langle {\vector S} \rangle}_{\xi i} 
       \equiv 
           \sum_{\vR_j \in v_{\xi i}} {\vector S}_j \, \, 
         / \sum_{\vR_j \in v_{\xi i}} 1 
     }$. 
Thus, we obtain 
\Equation{eq:HKinSiapprox}
{
     \begin{split}
     H_{\it K} \approx & 
               \int d^3 \vr' \sum_{i} 
                   {\hat J}_{\it K}(\vR_i - \vr') \, 
             \\
             & \times 
               {\langle {\vector S} \rangle}_{\xi i}
                 \cdot 
               \sum_{\sigma  \sigma'}
               \bigl [ 
                 \psi_{\sigma}^{\dagger}(\vr') 
                   {\boldsymbol{\sigma}}_{\sigma \sigma'} 
                 \psi_{\sigma'}(\vr') 
               \bigr ] . 
     \end{split}
     }
In actuality, the microscopic fluctuations omitted here 
may affect the superconductivity through the self-energy of the electrons. 
We regard those corrections to the normal state 
as being included in the effective mass $m^{*}$, if it exists. 
Since $\xi \gg a$, there are many localized spins 
in the volume $v_{\xi i}$, 
the antiferromagnetic components of the spins vanish on average, 
while the ferromagnetic moment remains. 
Therefore, we obtain 
${
     {\langle {\vector S} \rangle}_{\xi i}
       \approx ({\bar S} \sin \theta, 0, 0) 
     }$. 
For specific lattice structures, this equation is satisfied exactly. 
Furthermore, we assume that 
the localized spins on $z_{\it K}$ lattice sites around $\vr'$ take part in 
the Kondo interactions with the conduction electron at $\vr'$, 
and that all of these coupling constants are equal to $J_{\it K}$. 
Therefore, we obtain 
\Equation{eq:HKapproxresult}
{
     H_{\it K} \approx 
               h_{\it ex} 
               \int d^3 \vr' 
               \sum_{\sigma  \sigma'}
               \bigl [ 
                 \psi_{\sigma}^{\dagger}(\vr') 
                   {\sigma}^{x}_{\sigma \sigma'} 
                 \psi_{\sigma'}(\vr') 
               \bigr ] , 
     }
where $h_{\it ex}$ denotes the exchange field 
$h_{\it ex} = z_{\it K} J_{\it K} {\bar S} \sin \theta$. 
From \eq.{eq:sinthetaH}, we obtain 
\Equation{eq:hex}
{
     h_{\it ex} = 
       \frac{g_{\it s} \mu_{\it B}}{2} 
       \frac{z_{\it K} J_{\it K}}{z J} H , 
     }
for $H \leq H_{\it AF}$. 
For $H \geq H_{\it AF}$, 
we have $h_{\it ex} = z_{\it K} J_{\it K} {\bar S}$.

Taking into account the exchange field $h_{\it ex}$ 
and the external field $h$, 
the kinetic energy term of the conduction electron Hamiltonian 
is written as 
\Equation{eq:H0}
{
     \begin{split}
     H_0 
     & = 
          \sum_{\sigma \sigma'} \int d^3 \vr \, \, 
             \psi_{\sigma}^{\dagger}(\vr) \, 
             \Bigl [ 
             \frac{-\hbar^2}{2m^{*}} 
               \bigl (
                 {\boldsymbol \nabla} - \frac{e}{c} {\vector A}(\vr)
               \bigr )^2 
               \delta_{\sigma \sigma'}
     \\
     & \hsp{4} 
             - 
             (1 - \delta_{\sigma \sigma'}) \, 
             {\tilde h} \, 
             \Bigr ]
             \psi_{\sigma'}(\vr) , 
     \end{split}
     }
where $m^{*}$ and ${\tilde h}$ denote the band effective mass 
and the renormalized Zeeman energy $\tilde h = h - h_{\it ex}$, 
respectively. 
Here, we have defined $h = \mu_{\it e} H$ 
and $\mu_{\it e} = g_{\it e} \mu_{\it B}/2$ 
where $g_{\it e}$ denotes the $g$-factor of the conduction electrons. 
For $H \leq H_{\it AF}$, we have 
\Equation{renZeeman}
{
     \tilde h 
        = \bigl ( 
            1 - \frac{g_{\it s} z_{\it K} J_{\it K}}{g_{\it e} z J}
          \bigr ) \, h 
        = \bigl ( 
            1 - \frac{J_{\it K}}{J_{\it AF}'}
          \bigr ) \, h 
     }
obtained by \eq.{eq:hex}, 
where we have defined 
${
     J_{\it AF}' \equiv g_{\it e} z J / g_{\it s} z_{\it K} 
     }$. 
In contrast, for $H \geq H_{\it AF}$, 
we have ${\tilde h} = h - z_{\it K} J_{\it K} {\bar S}$

In \eq.{renZeeman}, it is explicit that the Zeeman energy is reduced 
by the exchange field when $J_{\it K} > 0$. 
It is found that 
the reduction effect becomes maximum when $J_{\it K} = J_{\it AF}'$. 
Obviously, this reduction mechanism is not affected 
by the anisotropy of the order parameter. 
Hence, we consider the $s$-wave superconductor as an example.

Applying the standard theory of superconductivity 
to the present system, 
we obtain the integral equation 
for the superconducting transition 
temperature $T_{\it c}$~\cite{Hel64,Sch80,Luk87,Shi96,Shi97a} 
\Equation{Tcequation}
{
     \begin{split}
     \ln \frac{T_{\it c}}{T_{\it c}^{(0)}} 
     & = \frac{1}{2} \int_0^{\infty} d x \frac{1}{\sinh x}
       \int_0^{\pi} d \theta \sin \theta 
     \\
     & \hsp{2} 
       \times 
       \Bigl [
         \exp 
           \bigl ( - \frac{\kappa'}{4} x^2 \sin^2 \theta \bigr ) 
         \cos ( 2 {\tilde h}' x ) 
       - 1
       \Bigr ] , 
     \end{split}
     }
where 
$T_{\it c}^{(0)}$ denotes the zero-field transition temperature, 
and we have defined 
${
     \kappa' \equiv ( v_{\it F}/2 \pi \Tc)^2 2 |e| H /c
     }$
and 
${
     {\tilde h}' \equiv {\tilde h}/2 \pi \Tc
     }$. 
Here, we have ignored the possibility of 
the Fulde-Ferrell-Larkin-Ovchinnikov state. 
The extension to include it is straightforward~\cite{Shi03,Shi97a}. 
From the form of $\kappa'$, it is convenient to define a constant 
\Equation{adef}
{
     a_{\it m} \equiv 
         \bigl ( 
           \frac{v_{\it F}}{2 \pi T_{\it c}^{(0)}} 
         \bigr )^2 
         \frac{2 |e|}{c} , 
     }
with which we can write 
${
     \kappa' = a_{\it m} H ( T_{\it c}^{(0)}/\Tc )^2 
     }$. 
We also define the ratio of 
the scales of the paramagnetic and orbital effects as 
\Equation{rmdef}
{
     r_{\it m} \equiv 
         \frac{h/2 \pi T_{\it c}^{(0)}}{ a_{\it m} H } 
         = 
         \frac{g_{\it e} \pi T_{\it c}^{(0)}}{8 \epsilon_{\it F}}
         \frac{m^{*}}{m} , 
     }
where $\epsilon_{\it F} \equiv m^{*} v_{\it F}^2/2$.

\begin{figure}
\vspace{8ex} 
\includegraphics[width=6.5cm]{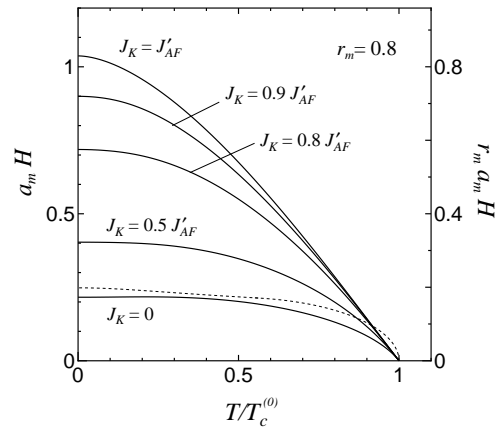}
\caption{
Temperature dependence of upper critical fields 
for $r_{\it m} = 0.8$. 
The dotted curve shows the pure Pauli paramagnetic limit~\cite{HPnote} 
in the absence of the orbital pair-breaking effect at $J_{\it K} = 0$, 
where $g_{\it e} = 2$. 
The curves of the upper critical field for $J_{\it K}/J_{\it AF}' = 1.1$, 
1.2, and 1.5 coincide with 
those for $J_{\it K}/J_{\it AF}' = 0.9$, 0.8, and 0.5, respectively. 
} 
\label{fig:Hc2rm08}
\end{figure}

\begin{figure}
\vspace{8ex}
\includegraphics[width=6.5cm]{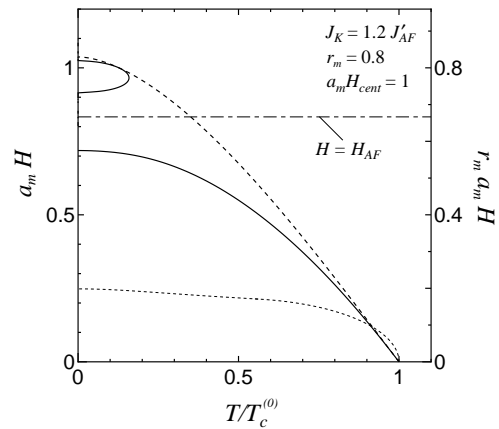}
\caption{
Critical fields of superconductivity 
for $J_{\it K} = 1.2 J'_{\it AF}$, 
$r_{\it m} = 0.8$ and $a_{\it m} H_{\it cent} = 1$. 
The dotted and short dashed curves show 
the pure paramagnetic and pure orbital limits, respectively. 
The dot-dashed line shows 
the critical field of the antiferromagnetic phase. 
}
\label{fig:Hc2rm08FISC}
\end{figure}

Equation~\refeq{Tcequation} is solved numerically. 
Figure~\ref{fig:Hc2rm08} shows the results 
for $r_{\it m} = 0.8$ and $J_{\it K} < J'_{\it AF}$. 
It is found that the upper critical field is enhanced as 
$J_{\it K}$ increases up to $J_{\it AF}'$. 
When $J_{\it K} \ne 0$, the upper critical field $H_{\it c2}$ 
can exceed the Pauli paramagnetic limit $H_{\it P}(T)$. 
In particular, for $J_{\it K} \sim J_{\it AF}'$, 
the upper critical field $H_{\it c2}$ can reach fourfold 
the Pauli limit $H_{\it P}$. 
When $J_{\it K} = J_{\it AF}'$, the pure orbital limit $H_{\it c20}(T)$ 
is recovered. 
For larger $r_{\it m}$, {\it i.e.}, weaker orbital effect, 
the ratio $H_{\it c2}/H_{\it P}$ increases.

If there is a temperature region in which $H_{\it c2}(T) > H_{\it AF}$ 
in Fig.~\ref{fig:Hc2rm08}, 
the curve of $H_{\it c2}(T)$ should be modified there 
so that it saturates more rapidly, 
because the exchange field is constant for $H > H_{\it AF}$. 
In the scale of the right vertical axis, 
$H = H_{\it AF}$ gives 
$r_{\it m} a_{\it m} H_{\it AF} 
= g_{\it e} zJ{\bar S}/2 \pi g_{\it s} T_{\it c}^{(0)}$. 
Hence, if $g_{\it e}/g_{\it s} \gg 1$ 
or if $zJ{\bar S}/T_{\it c}^{(0)} \sim T_{\it AF}/T_{\it c}^{(0)} \gg 1$, 
the upper critical field $H_{\it c2}(T)$ does not exceed $H_{\it AF}$ 
in the whole temperature region.

In a high-field region in which $H \geq H_{\it AF}$, 
where the spins are aligned uniformly, 
the mechanism is reduced to the Jaccarino-Peter mechanism. 
In this case, the Zeeman energy completely vanishes 
at $H = z_{\it K} J_{\it K} {\bar S}/\mu_{e} \equiv H_{\it cent}$. 
Therefore, if $H_{\it cent} + H_{\it P} > H_{\it AF}$ is satisfied, 
we need to examine the possibility of FISC. 
Since $H_{\it cent} / H_{\it AF} = J_{\it K} / J'_{\it AF}$, 
FISC occurs only when $J_{\it K} \gsim J_{\it AF}'$~\cite{Shi02b}. 
Figure~\ref{fig:Hc2rm08FISC} shows the critical fields 
for $J_{\it K} = 1.2 J'_{\it AF}$, 
$r_{\it m} = 0.8$ and $a_{\it m} H_{\it cent} = 1$. 
This set of the parameter values gives 
$a_{\it m} H_{\it AF} \approx 0.83$, 
and if $g_{\it s} = g_{\it e} = 2$, $T_{\it AF} \sim z J {\bar S} 
  = 2 \pi g_{\it s} T_{\it c}^{(0)} r_{\it m} a_{\it m} H_{\it AF}/ g_{\it e} 
  \sim 4 \, T_{\it c}^{(0)}$. 
In Fig.~\ref{fig:Hc2rm08FISC}, 
we find an area of FISC 
inside the region of $H_{\it c20}(T) > H > H_{\it AF}$.

In this context, the compound ${\rm CePb_3}$ corresponds to 
the case of $H_{\it cent} \gg H_{\it AF}$, 
{\it i.e.}, $J_{\it K} \gg J_{\it AF}'$, 
and $r_{\it m} \gg 1$~\cite{NoteCePb3}. 
In contrast, the compound ${\rm CePt_3Si}$ has two possibilities, 
{\it i.e.}, $J_{\it K} \gsim J_{\it AF}'$ and $J_{\it K} \lsim J_{\it AF}'$. 
For the former case, FISC might be observed 
at very high fields, 
if $H_{\it c20}(0) \gsim H_{\it cent} \gsim H_{\it AF}$~\cite{NoteCePt3Si}. 
In the compound ${\rm UPd_2Al_3}$, 
it was observed that $T_{\it AF} = 14.3$\,K 
and $T_{\it c}^{(0)} = 2.0$\,K~\cite{Gei91}. 
Such a high $T_{\it AF}$ suggests small $J_{\it K}/J_{\it AF}'$, 
so that the reduction of the paramagnetic effect is not pronounced. 
This coincides with the experimental fact that $H_{\it c2} < H_{\it P}$ 
and the absence of FISC in the compound ${\rm UPd_2Al_3}$. 
Generally speaking, the difference in the experimental results 
$T_{\it AF} \approx 1.1$\,K, 2.2\,K, and 14.3\,K in 
${\rm CePb_3}$, ${\rm CePt_3Si}$, and ${\rm UPd_2Al_3}$, respectively, 
is consistent in the present theory with their phase diagrams, 
if their $J_{\it K}$'s are of the same order.

Here, the condition $r_{\it m} \sim 1$ can be satisfied 
in heavy-fermion superconductors, but not in conventional metals, 
because 
$r_{\it m} \propto g_{\it e} T_{\it c}^{(0)}/m \vF^2 
= g_{\it e} (T_{\it c}^{(0)}/\epsilon_{\it F}) (m^{*}/m)$. 
Also in the scenario with equal spin pairing, 
we need $r_{\it m}$ of order 1, in order to reproduce 
a large ratio of $H_{\it c2}/H_{\it P}$ as observed.

In conclusion, we have shown that 
the Pauli paramagnetic pair-breaking effect can be 
considerably reduced in antiferromagnetic superconductors, 
even in the presence of the orbital pair-breaking effect, 
unless $r_{\it m} \ll 1$. 
One of the necessary conditions for the occurrence of the present mechansim 
is that the Kondo exchange coupling constant $J_{\it K}$ is positive 
and comparable to the scale of 
the antiferromagnetic exchange energy $J_{\it AF}'$. 
The present result may explain the high $H_{\it c2}$ 
observed in the compound ${\rm CePt_3Si}$. 
The phase diagrams of the compounds ${\rm CePt_3Si}$ 
and ${\rm CePb_3}$ can be understood 
within a unified framework. 
The positive $J_{\it K}$ can be attributed to superexchange 
or kinetic exchange processes. 
Hence, the condition $J_{\it K} \sim J \sim J_{\it AF}'$ can be satisfied, 
because the antiferromagnetic exchange interaction $J$ originates 
from exchange processes 
similar to those for $J_{\it K}$ in the same crystal structure. 
The hybrid ruthenate-cuprate compound \RSGCO~\cite{Pic99,Shi00,Nak02} 
is another possible candidate of the present mechanism, 
if one applies a parallel magnetic field to it. 
The antiferromagnetic long-range order with weak ferromagnetism 
due to the canted spin structure has been observed 
in this compound at zero field~\cite{Nak02}.

The author wishes to thank \mbox{Y.~Matsuda} and 
\mbox{T.~Oguchi} for useful discussions, 
\mbox{K.~V.~Samokhin} for useful comments, 
and \mbox{N.~Metoki} for informing us of their paper. 
This work was partly supported by 
a Grant-in-Aid for COE Research, No.\hspace{0.25ex}13CE2002, 
and a Grant-in-Aid for Scientific Research (C), No.\hspace{0.25ex}16540320, 
of the Ministry of Education, 
Culture, Sports, Science and Technology of Japan.




\begin{thebibliography}{99}
\bibitem{Cha62}
  B.~S.~Chandrasekhar: App. Phys. Lett. {\bf 1} (1962) 7. 
\bibitem{Clo62}
  A.~M.~Clogston: Phys. Rev. Lett. {\bf 9} (1962) 266. 
\bibitem{Shi02b} 
  H.~Shimahara: J. Phys. Soc. Jpn. {\bf 71} (2002) 713. 
\bibitem{Shi03} 
  H.~Shimahara: Physica B {\bf 329-333} (2003) 1442. 
\bibitem{Jac62} 
  V.~Jaccarino and M.~Peter: Phys. Rev. Lett. {\bf 9} (1962) 290. 
\bibitem{Meu84}
  H.~W.~Meul {\it et al.}: Phys. Rev. Lett. {\bf 53} (1984) 497. 
\bibitem{Lin85}
  C.~L.~Lin {\it et al.}: Phys. Rev. Lett. {\bf 54} (1985) 2541. 
\bibitem{Uji01}
  S.~Uji {\it et al.}: 
  Nature {\bf 410} (2001) 908. 
\bibitem{Cep02}  
  O.~C${\acute {\it e}}$pas, R.~H.~McKenzie and J.~Merino: 
  Phys. Rev. B {\bf 65} (2002) 100502R. 
\bibitem{Shi02a} 
  H.~Shimahara: J. Phys. Soc. Jpn. {\bf 71} (2002) 1644. 
\bibitem{Hou02}  
  M.~Houzet, A.~Buzdin, L.~Bulaevskii and M.~Maley: 
  Phys. Rev. Lett. {\bf 88} (2002) 227001. 
\bibitem{Bau04} 
  E.~Bauer {\it et al.}: Phys. Rev. Lett. {\bf 92} (2004) 027003. 
\bibitem{Fri04} 
  P.~A.~Frigeri, D.~F.~Agterberg, A.~Koga and M.~Sigrist: 
  Phys. Rev. Lett. {\bf 92} (2004) 097001. 
\bibitem{Sam04} 
  K.~V.~Samokhin, E.~S.~Zijlstra and S.~K.~Bose: 
  Phys. Rev. B {\bf 69} (2004) 094514. 
\bibitem{Sam04Note} 
  According to the theory in ref.~\cite{Sam04}, 
  the electron states are nondegenerate 
  almost everywhere in the momentum space 
  because of strong spin-orbit coupling, 
  and Cooper pairing occurs between electrons 
  with $(\vk,n)$ and $(-\vk,n)$, 
  which have the same energy $\epsilon_n(\vk) = \epsilon_n(-\vk)$, 
  where $n$ denotes the band index. 
  Since the states with $(\vk,n)$ and $(-\vk,n)$ are connected by 
  the time reverse transformation, 
  they cannot be equal spin states. 
  If the state with $(\vk,n)$ is up-spin rich, 
  the state with $(-\vk,n)$ must be down-spin rich. 
  Thus, the pair of the electrons with $(\vk,n)$ and $(-\vk,n)$ 
  experiences the Pauli paramagnetic pair-breaking effect. 
  In ref.~\cite{Sam04a}, it is shown that 
  the paramagnetic suppression of superconductivity is nonzero in all 
  directions in general, and the Knight shift has an unusual anisotropy of 
  the temperature dependence.
\bibitem{Met04}
  N.~Metoki {\it et al.}: 
  J. Phys. Condens. Matter {\bf 16} (2004) L207. 
\bibitem{NoteHs} 
  We define the coupling constant $J_{ij}$ as one including 
  the corrections from the conduction electrons. 
  In actuality, the influence of the conduction electron 
  may not be included completely only through such corrections, 
  because various spin structures other than the canted structure 
  can occur~\cite{Ham95}. 
  However, we neglect such possibilities phenomenologically, 
  and consider only the cases in which the magnetic subsystem is 
  effectively described by \eq.{eq:Hs}. 
\bibitem{Hel64}
  E.~Helfand and N.~R.~Werthamer: 
  Phys. Rev. Lett. {\bf 13} (1964) 686. 
\bibitem{Sch80}
  K.~Scharnberg and R.~A.~Klemm: 
  Phys. Rev. B {\bf 22} (1980) 5233. 
\bibitem{Luk87}
  I.~A.~Luk'yanchuk and V.~P.~Mineev: 
  Sov. Phys. JETP {\bf 66} (1987) 1168. 
\bibitem{Shi96}
  H.~Shimahara, S.~Matsuo and K.~Nagai: 
  Phys. Rev. B {\bf 53} (1996) 12284. 
\bibitem{Shi97a}
  H.~Shimahara and D.~Rainer: 
  J. Phys. Soc. Jpn. {\bf 66} (1997) 3591. 
\bibitem{HPnote}
  In the present scale, 
  the Pauli limit $H_{\it P} = \Delta_0/\sqrt{2}\mu_{\it B}$ at $T = 0$ 
  corresponds to 
  $a_{\it m} H_{\it P} = h_{\it P} /2 r_{\it m} \pi T_{\it c}^{(0)} 
  = g_{\it e} /4 \sqrt{2} e^{\gamma} r_{\it m}$, 
  where $\gamma = 0.57721 \cdots$. 
  Therefore, we obtain $a_{\it m} H_{\it P} \approx 0.2/r_{\it m}$, 
  if $g_{\it e} \approx 2$. 
\bibitem{NoteCePb3}
  It was observed that 
  $H_{\it cent} \gsim 14$\,T and $H_{\it AF} \approx 5$\,T~\cite{Lin85}, 
  which give $J_{\it K} \gg J_{\it AF}'$. 
  Furthermore, since the FISC was observed for $H \gsim 14$\,T, 
  $H_{\it c20}(0) \gg 14$\,T must be satisfied, 
  while $H_{\it P} \gsim 1.86 \times 0.6 \sim 1.1$\,T, 
  since $T_{\it c}^{(0)} \gsim 0.6$\,K. 
  Hence, we obtain $r_{\it m} \sim 0.2 H_{\it c20}/H_{\it P} \gg 3 \gg 1$. 
\bibitem{NoteCePt3Si}
  If $J_{\it K}$'s are on the same order 
  in the compounds ${\rm CePb_3}$ and ${\rm CePt_3Si}$, 
  it is expected that $H_{\it cent} \gsim 14$\,T 
  in the compound ${\rm CePt_3Si}$. 
  Furthermore, comparing the values of 
  $T_{\it c}^{(0)}$ and $T_{\it AF}^{(0)}$ of the two compounds, 
  it seems plausible that 
  $H_{\it c20}(0)$ should be much larger than 14\,T, 
  and that $H_{\it AF} \sim 10$\,T. 
  In this case, 
  the condition $H_{\it c20}(0) > H_{\it cent} > H_{\it AF}$ is satisfied. 
\bibitem{Gei91}
  C.~Geibel {\it et al.}: Z. Phys. B {\bf 84} (1991) 1. 
\bibitem{Pic99}
  W.~E.~Pickett, R.~Weht and A.~B.~Shick: 
  Phys. Rev. Lett. {\bf 83} (1999) 3713. 
\bibitem{Shi00}
  H.~Shimahara and S.~Hata: 
  Phys. Rev. B {\bf 62} (2000) 14541. 
\bibitem{Nak02} 
  K.~Nakamura and A.~J.~Freeman: 
  Phys. Rev. B {\bf 66} (2002) 140405R. 
\bibitem{Sam04a}
  K.~V.~Samokhin: cond-mat/0405447; cond-mat/0404407. 
\bibitem{Ham95}
  M.~Hamada and H.~Shimahara: 
  Phys. Rev. B {\bf 51} (1995) 3027; 
  J. Phys. Soc. Jpn. {\bf 65} (1996) 552. 
\end{thebibliography}
\end{document}